
\input harvmac
\overfullrule=0pt

\Title{HUTP-92/A046}{Rogers Dilogarithm in Integrable Systems*
\footnote{}{\hskip -0.14truein *~ To appear in the proceedings of the XXI
Differential Geometry
 Methods in Theoretical Physics, Tianjin, China 5-9 June 1992.  Preprint
distribution partially supported by NSF Grant PHY~87-14654, and a Packard
Fellowship.}}

\centerline{Atsuo Kuniba}
\bigskip\centerline{Department of Mathematics, Kyushu University}
\centerline{Fukuoka 812 JAPAN}
\vskip0.3cm
\centerline{and}
\vskip0.3cm
\centerline{Tomoki Nakanishi\footnote{$^\dagger$}
{Permanent Address: Department of Mathematics, Nagoya University,
Nagoya 464 Japan}}
\bigskip\centerline{Lyman Laboratory of Physics, Harvard University}
\centerline{Cambridge, MA 02138 USA}


\vskip .3in
We discuss some curious aspects of the Rogers dilogarithm appearing
in integrable systems in two dimensions.

\Date{09/92} 


\newsec{Introduction}
\indent
This note is a brief exposition of the appearance of the
Rogers dilogarithm function in relation to integrable
lattice models and conformal field theory (CFT) in two dimensions.
The content is mainly the known facts in refs.${}^{1,2}$ but also includes
a few new informations based on a collaboration with Junji Suzuki.
\par
The Rogers dilogarithm is a function of a variable $x$ defined by
$$L(x) = -{1 \over 2}\int_0^x \bigl(
{log (1-y) \over y} + {log y \over 1-y} \bigr) dy \quad (0 \le x \le 1).
\eqno(1)
$$
The following identity is know essentially due to refs.${}^3$
$$
{6 \over \pi^2} \sum_{m=1}^\ell L\bigl(
{sin^2 {\pi \over \ell + 2} \over sin^2 {\pi(m+1) \over \ell + 2}}
\bigr) = {3\ell \over \ell + 2},
\quad\hbox{ for } \ell \in {\bf Z}_{\ge 1}.\eqno(2)
$$
In the above, the rhs is the well known value of the central charge
for the level $\ell \,\, A^{(1)}_1$ WZW model in conformal field theory.
On the other hand, the argument of the dilogarithm
has the form $(Q_m)^{-2}$,
where $Q_m$ is the $m+1$ dimensional
irreducible $A_1$ character specialized to some
``rational point".
Thus Eq.\ 2 is connecting the two important quantities in
the $A_1$-related theory, i.e.,
the central charge and the specialized character.
\par
In fact, there is a conjectural generalization of the identity Eq.\ 2
into arbitrary classical simple Lie algebra $X_r^{4,1}$.
$$
{6 \over \pi^2} \sum_{a=1}^r \sum_{m=1}^{t_a \ell} L\bigl(
f^{(a)}_m
\bigr) = {\ell dim X_r  \over \ell + g},
\quad\hbox{ for } \ell \in {\bf Z}_{\ge 1}.\eqno(3)
$$
Here, $g$ denotes the dual Coxeter number and
$t_a$ is the integer defined as the ratio of the $a-$th Kac
and dual Kac label.
The argument $0 \le f^{(a)}_m \le 1$ arises through
thermodynamic Bethe ansatz (TBA) analysis.
Here we shall give its definition only for
$X_r = A_r$. (See refs.${}^{1,4,5}$ for the general case.)
$$
f^{(a)}_m = f^{(a)}_m(z=0), \qquad
f^{(a)}_m(z) = 1 - {Q^{(a)}_{m+1}(z)Q^{(a)}_{m-1}(z)
\over Q^{(a)}_m(z){}^2}, \eqno(4)
$$
where $Q^{(a)}_m(z)$ is the irreducible $A_r$ character
with the highest weight $m\Lambda_a$.
($z \in $ dual space of the Cartan subalgebra and
$\Lambda_a$ is the $a-$th fundamental weight.
Nodes on the Dynkin diagram are enumerated according to ref.${}^1$.)
In general, the quantity $Q^{(a)}_m(z)$ is
a Yangian character and we adopt the $z$ dependence as given in
ref.${}^2$.
The conjecture (4) was firstly systematically used in the
TBA analysis in refs.${}^{6,1}$.

\newsec{Scaling Dimensions from Dilogarithm}
\indent
There is a generalization${}^2$ of Eq.\ 3 so as to include
the parafermion scaling dimensions (modulo  integer) in CFT
$\Delta^{\Lambda}_{\lambda} = {(\Lambda \vert \Lambda + 2\rho) \over
2(\ell + g)} - {\vert \lambda \vert^2 \over 2g}$,
where $\Lambda$ is a level $\ell$ dominant integral weight of
$X^{(1)}_r$ and $\lambda \in \Lambda + $ root lattice.
This is achieved by considering the specialization
$ f^{(a)}_m(z=\Lambda)$ instead of the principal one in Eq.\ 4.
Now that the $ f^{(a)}_m(\Lambda)$ is a
complex number in general, one can consider
various analytic continuations $L_{a,m}(x)$ of $L(x)$.
Leaving all the technical points${}^2$, we have a conjecture
$$
{6 \over \pi^2} \sum_{a=1}^r \sum_{m=1}^{t_a \ell-1}
L_{a,m}\bigl(f^{(a)}_m(\Lambda)\bigr)
+ \hbox{ Logarithmic terms} = {\ell dim X_r  \over \ell + g} - r
 -24 (\Delta^{\Lambda}_{\lambda} + \hbox{ integer}). \eqno(5)
$$
Here, $\lambda$
depends on the integration contour along which
the $L(x)$ is analytically continued.
Its explicit form and the logarithmic terms can be found in ref.${}^2$.
See also refs.\ ${}^{7,8,9,10}$ for some physical aspects.
Here we shall only present a conjecture
on the value $Q^{(a)}_{t_a\ell}(\Lambda)$,
under which the congruence $\lambda \equiv \Lambda$ (modulo  root lattice)
can be verified directly.
Put $k =$ number of the Kac labels $a_i \, (0 \le i \le r)$ equal to 1.
Then we conjecture that $Q^{(a)}_{t_a\ell}(\Lambda)$ is a
$k-$th root of unity as
$$\eqalign{
Q^{(a)}_{t_a\ell}(\Lambda) &=
\hbox{exp}(-2\pi i c(\Lambda){\overline \gamma}_a/k)
\quad \hbox { for } X_r \neq D_r,\cr
&= \hbox{exp}(2\pi i c_2(\Lambda)\gamma^{(2)}_a/k)
\quad \hbox { for } X_r =  D_r, \, r = \hbox{odd},\cr
&= \hbox{exp}(2\pi i ((c_2(\Lambda)-rc_1(\Lambda))\gamma^{(1)}_a
                    + c_1(\Lambda)\gamma^{(2)}_a)/k)
\quad \hbox { for } X_r = D_r,\, r = \hbox{even}.\cr
}\eqno(6)$$
Here, for $\Lambda = \sum_{a=1}^r \mu_a \Lambda_a$,  we have set
$c(\Lambda)
= \sum_a \gamma_a \mu_a \,  \hbox{ mod }\, k \,
\hbox{ for } X_r \neq D_r$ and
$c_i(\Lambda)
= \sum_a \gamma_a^{(i)} \mu_a \,\, \hbox{ mod }\,  k/(3-i)
\, \hbox{ for } X_r = D_r, \,
i = 1,2$
and
$\gamma$ is the rank-dimensional integer vector given by
$$
\eqalign{
&A_r, B_r, C_r, E_6: \gamma_a = a,\cr
&D_r: \gamma^{(1)}=(0,\ldots,0,1,1), \quad
\gamma^{(2)} = (2,4,6,\ldots,2(r-2),r-2,r),\cr
&E_7: \gamma = (0,0,0,1,0,1,1),\cr
&E_8,F_4,G_2: \gamma = (0,\ldots,0).\cr
}\eqno(7)$$
Finally, ${\overline \gamma} = -\gamma$ if $X_r = A_r$ and
${\overline \gamma} = \gamma$ of the dual algebra of  $X_r$ if
$X_r \neq A_r, D_r$.
Eq.\ 6 is a special solution of
$\prod_b Q^{(b)}_{t_b \ell}(\Lambda)^{C_{a b}}=1$
where $C$ is the Cartan matrix of $X_r$.
{}From numerical tests it also seems valid that
$Q^{(a)}_m(\Lambda) = Q^{(a)}_{t_a\ell}(\Lambda)
Q^{(a)}_{t_a\ell - m}(\Lambda){}^{\ast}$ for
$-1 \le m \le t_a\ell + 1$ and
$Q^{(a)}_{t_a\ell + 1}(\Lambda) = 0$
for any level $\ell$ dominant
integral weight $\Lambda$.
These are interesting arithmetic properties of
the specialized Yangian characters.
We note especially that $Q^{(a)}_m(0)$ appears${}^1$
as the high temperature limit
of $log ( entropy )$ per site in the TBA system connected to $X_r$.
This implies that $Q^{(a)}_m(0)$ yields the largest eigenvalue
of the incidence matrix for a fusion $X^{(1)}_r$ RSOS model.

\newsec{Functional Relations}
\indent
The Yangian character  $Q^{(a)}_m(z)$ is known to satisfy
interesting recursion relations${}^{4,5}$.
For example in $X_r = A_r$ case,
$$
Q^{(a)}_m(z){}^2 = Q^{(a)}_{m+1}(z)Q^{(a)}_{m-1}(z) +
                   Q^{(a+1)}_m(z)Q^{(a-1)}_m(z)\eqno(8)
$$
and similar relations are known for all the other algebras.
Furthermore there is a ``spectral parameter dependent version
(or Yang-Baxterization)" of these relations.
Below we shall describe it briefly for $X_r = A_r$.
In ref.${}^6$ Bazhanov and Reshetikhin wrote down a system of
functional relations among the row to row transfer matrices
for the fusion $A^{(1)}_r$ model${}^{11}$.
$$\eqalignno{
T^{\xi, \eta}(u) &= \hbox{det}\, \bigl(
T^{\xi, (\eta_i-i+j)\Lambda_1}(u+\eta_1+i-\eta_i-1)
\bigr)_{1 \le i,j \le \eta^{\prime}_1} &(9\hbox{a})\cr
&= \hbox{det}\, \bigl(
T^{\xi, \Lambda_{\eta^{\prime}_i-i+j}}(u+\eta_1-j)
\bigr)_{1 \le i,j \le \eta_1}\, , &(9\hbox{b})\cr
}$$
where $\xi$ (resp. $\eta$) is the Young diagram representing
the fusion type in the horizontal (resp. vertical)
direction and the $T^{\xi, \eta}(u)$ is
transferring the states into the vertical direction.
$\eta^{\prime} = [\eta^{\prime}_1, \ldots , \eta^{\prime}_{\eta_1}]$
is the transpose of
$\eta = [\eta_1, \ldots , \eta_{\eta^{\prime}_1}]$
and $u$ is the spectral parameter entering the solution of the
Yang-Baxter equation that underlies the model${}^{11}$.
Eq.\ 9 is a ``quantum analogue" of the
2nd Weyl character formula (Jacobi-Trudi's formula)${}^6$.
(In fact Eq.\ 9 is different from that in ref.${}^6$ and the
alternation is based on a private communication with V.V. Bazhanov.
Note that $T^{\xi, \eta}(u)$ may be regarded as representing an eigenvalue
thanks to the commutatibity.)
Now let $\eta$ be the $a$ by $m$ rectangular Young diagram
corresponding to $m\Lambda_a$ and write the $T^{\xi, \eta}(u)$
as $T^{(a)}_m(u)$.
Then from Eq.\ 9 one can prove the functional relation
$$
T^{(a)}_m(u)T^{(a)}_m(u+1) = T^{(a)}_{m+1}(u) T^{(a)}_{m-1}(u+1) +
                   T^{(a+1)}_m(u) T^{(a-1)}_m(u+1),\eqno(10)
$$
which is a ``Yang-Baxterization" of Eq.\ 8.
{}From this we find that the combination
$$y^{(a)}_m(u+{a+m \over 2}) =
{T^{(a)}_{m+1}(u)T^{(a)}_{m-1}(u+1) \over
T^{(a+1)}_m(u)T^{(a-1)}_m(u+1)} \eqno(11)
$$
solves essentially the following
$U_q(A^{(1)}_r)$ functional relation proposed in ref.${}^2$
$$
y^{(a)}_m(u+{1 \over 2})y^{(a)}_m(u-{1 \over 2})
= {(1 + y^{(a)}_{m+1}(u))(1 + y^{(a)}_{m-1}(u)) \over
   (1 + y^{(a+1)}_m(u)^{-1}) (1 + y^{(a-1)}_m(u)^{-1})}. \eqno(12)
$$
As noted in ref.${}^2$, the above equation is also satisfied by
$e^{\epsilon^{(a)}_m(u)}$ ($\epsilon^{(a)}_m(u)$: pseudo energy )
in the TBA at $\infty$ temperature.
In this way, one can roughly say that the TBA equation
has a solution in terms of
``Yang-Baxterized" Yangian characters.
It is interesting to note that Eq.\ 10 can be viewed as a simplest
example of the Pl\"ucker relation under Eq.\ 9.

\beginsection References

\item{1.}{A.Kuniba, ``Thermodynamics of the $U_q(X^{(1)}_r)$
Bethe ansatz system with $q$ a root of unity" , to appear in
Nucl.Phys.B.}
\item{2.}{A.Kuniba and T.Nakanishi,
``Spectra in conformal field theories from the Rogers dilogarithm",
ANU preprint (1992) SMS-042-92.}
\item{3.}{A.N.Kirillov and N.Yu.Reshetikhin,
J.Phys.{\bf A20} (1987) 1587.}
\item{4.}{A.N.Kirillov, Zap.Nauch.Semin.LOMI {\bf 164} (1987) 121.}
\item{5.}{A.N.Kirillov and N.Yu.Reshetikhin,
Zap.Nauch.Semin.LOMI {\bf 160} (1987) 211.}
\item{6.}{V.V.Bazhanov and N.Yu.Reshetikhin, J.Phys.{\bf A23} (1990)
1477.}
\item{7.}{M.J.Martins, Phys.Rev.Lett. {\bf 67} (1991) 419.}
\item{8.}{T.R.Klassen and E.Melzer, Nucl.Phys.{\bf B370} (1992) 511.}
\item{9.}{P.Fendley, Nucl.Phys. {\bf B374} (1992) 667.}
\item{10.}{A.Kl\"umper and P.A.Pearce, Physica {\bf A183} (1992) 304.}
\item{11.}{M.Jimbo, A.Kuniba, T.Miwa and M.Okado, Commun.Math.Phys.
{\bf 119} (1988) 543.}
\bye